\documentclass[prb,aps,amsfonts]{revtex4}
\usepackage{amsmath}
\usepackage{epsfig}

\newcommand\algo[1]{{\bf #1}}                                                       
\begin{document}

\title{Electrostatics in Periodic Slab Geometries II}
\author{Jason de Joannis}
\email{joannis@mpip-mainz.mpg.de}
\author{Axel Arnold}
\email{arnolda@mpip-mainz.mpg.de}
\author{Christian Holm}
\email{holm@mpip-mainz.mpg.de}
\affiliation{Max-Planck-Institut f\"{u}r Polymerforschung,
Ackermannweg 10, 55128 Mainz, Germany}                 
                                                      
\date{\today}

\begin{abstract}
  In a previous paper a method was developed to subtract the interactions due
  to periodically replicated charges (or other long-range entities) in one
  spatial dimension. The method constitutes a generalized ``electrostatic
  layer correction'' (\algo{ELC}) which adapts any standard 3D summation
  method to slab-like conditions. Here the implementation of the layer
  correction is considered in detail for the standard Ewald (\algo{EW3DLC})
  and the \algo{P3M} mesh Ewald (\algo{P3MLC}) methods. In particular this
  method offers a strong control on the accuracy and an improved computational
  complexity of $O(N\log N)$ for mesh-based implementations. We derive
  anisotropic Ewald error formulas and give some fundamental guidelines for
  optimization. A demonstration of the accuracy, error formulas and
  computation times for typical systems is also presented.
\end{abstract}


\preprint{ELC II}
\maketitle


\section{ Introduction}

\label{intro}

Long-range forces, namely gravitation and electromagnetic interactions, are
difficult to treat exactly in many-particle systems. In condensed matter,
biological and solid state physics, there is considerable interest in the
simulation of charged and polar molecules embedded in a slab in which the
particles are replicated periodically in two directions. The applications
range widely to include soap bubbles, cell membranes and electrochemistry.
Surprisingly Ewald and other rapidly convergent methods for computing
long-range interactions in infinitely periodic systems are computationally
more demanding when the system is periodic in only one or two of the three
spatial dimensions because of the breaking of spatial symmetry.

Several ``two-dimensional Ewald'' (\algo{EW2D}) methods have already
been proposed\cite{widmann97a}.  The most successful is that
introduced by Parry\cite{parry75a,parry76a} and
others\cite{heyes77a,deleeuw82a}. This method can be used to obtain
accurate results but is limited to small systems (i.e. $10^{2}$ to
$10^{3}$ charges). This springs from the fact that the pair
separations, ${\bf r}_{ij}$, are no longer separable in the Fourier
summation because of the necessary decoupling of the aperiodic $z$
component from the periodic $x$ and $y$ components. Therefore it
appears that the complexity (scaling of time with number of charges)
of this method can be no better than quadratic. Also the one dimensional
Ewald method\cite{porto00a} suffers from the same problem.

A promising alternative to Ewald summation is possible using a
convergence-factor technique\cite{sperb01a}. The basic Coulomb pair
interaction is multiplied by one with the factor $\lim_{\beta
\rightarrow \infty }e^{-\beta r_{ijn}}$. It is crucial to prove that
the limit can be taken outside of the summation and now the modified
potential is more readily susceptible to analysis. In a recent
study,\cite{arnold01b} Arnold and Holm derive a two-dimensional
version of this method with complexity $N^{5/3}$ along with accurate
error formulas. This method is particularly attractive when extremely
high accuracy is desired. We will use this method, abbreviated as
\algo{MMM2D} and the \algo{EW2D} algorithm mentioned above as
reference methods.

\begin{figure}[tbp]
\begin{center}
\includegraphics[scale=0.5,angle=-90]{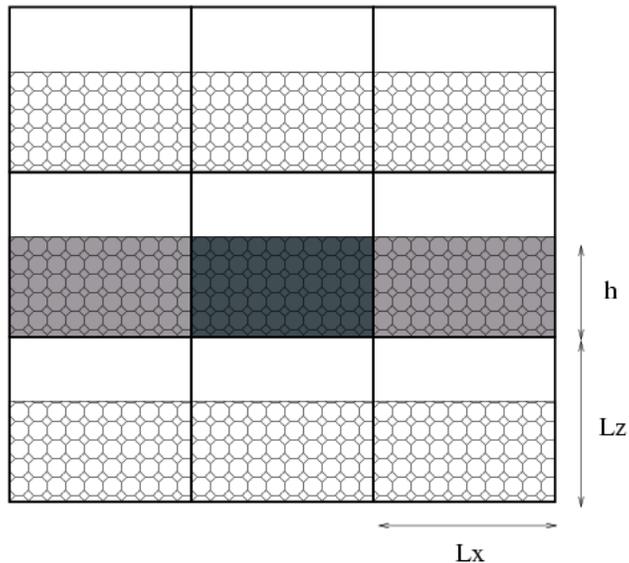}
\end{center}
\caption{Schematic of periodicity.}
\label{periodicity}
\end{figure}

One idea, used several times to study water interfaces, introduces a
spatial constraint within the simulation cell. If the box has
dimensions $L_{x}\times L_{y}\times L_{z}$, the particles may have
$z$-coordinates on $[0,h)$, while the space within $[h,L_{z})$ (the gap size)
remains empty \cite{empty-space}. The primary cell is, as usual,
replicated in all three directions periodically. The effect of this is
to create a ``primary layer'' and an empty layer followed by an
infinite array of intercalated image and empty layers. Figure
\ref{periodicity} summarizes the essential geometrical considerations
in this problem.
While this idea seems trivially to be correct, it has one flaw. The
flaw can be realized by attempting to reproduce the results of an
\algo{EW2D} calculation using a three dimensional Ewald summation
(\algo{EW3D}) as just described. One will find that the results always
differ by a dipole-moment dependent constant.  This fact, although at
first thought to be merely a slow convergence issue, was noticed by
Spohr \cite{spohr97a}. The \algo{EW3D} formula contains a shape
dependent dipole-term whose origin was mathematically proven in the
1980's \cite{deleeuw80a,deleeuw80b}, that reflects the naturally
chosen spherical order of summation in the conditionally convergent
Coulomb sum.  For the case of layers, however, it is necessary to use
a {\it slab-wise summation order}. This was realized by Yeh and
Berkowitz\cite{yeh99a}, who applied a theory due to E. R. Smith
\cite{smith81a} for infinite crystals of various shapes in order to
obtain the correct dipole term for a slab-wise summation order.  
Another complication
arises, if the periodic supersystem is surrounded by some medium with
a different dielectric constant. This medium has a polarizing effect
on the particles in the simulation cell no matter how large the
supersystem is. For the remainder of this paper we consider only the
simple case of zero contrast for which the dielectric constant is the
same inside and outside the supersytem.  This is often referred to as
the vacuum boundary condition.

One of the advantages of using this method from a practical point of
view is that any standard Ewald program may be used with only minor
modification. It is easy to show numerically that the use of this new
dipole term and sufficient spacing ($L_{z}-h$), yields the same result
as two-dimensional ``brute force'', \algo{MMM2D} and \algo{EW2D}
summation methods. As
explained in a previous paper, referred to here as ``paper
I''\cite{arnold01d}, errors introduced due to the image layer effects
decay exponentially with the size of the gap.  
This method is sometimes referred to as a corrected
three-dimensional Ewald sum (\algo{EW3DC})\cite{yeh99a}. 
Like the original Ewald
method it has a complexity of $N^{3/2}$ and the equations can be
discretized onto a mesh yielding an $N(\log N)^{3/2}$ complexity. In
paper I \cite{arnold01d}, we derived an analytic electrostatic layer
correction term, called \algo{ELC}-term, that subtracts the
interactions of the unwanted slab replicas. This term can be evaluated
linearly in $N$, and has full error control. The use of the \algo{ELC}
term and the change of summation order, shortly called the \algo{ELC}
method, enables us to adapt any 3D summation method, such as the
non-Ewald convergence-factor\cite{sperb01a} and multipole
expansion\cite{greengard87a} methods, to slab geometries. The latter
of these, though in possession of a slightly better scaling, has such
a considerable amount of overhead that it is useful only for much
larger systems than normally used\cite{esselink94a}.  The
abbreviations and complexities of some of the two-dimensional methods
available at present are summarized in Table \ref{tab1}, where the
ending LC always denotes the use of the \algo{ELC} method. In this
terminology the \algo{EW3D} algorithm plus the \algo{ELC} method is
called \algo{EW3DLC} and so on.

\begin{table}[tbp]
\caption{Methods for slab geometry.}
\begin{tabular}{c|c|c}
    method    &        authors                         & complexity \\ \hline
 \algo{EW2D}  & Parry; HBC; DP \cite{parry75a,heyes77a,deleeuw82a} & $N^{2}$ \\
 \algo{Lekner}& Lekner \cite{lekner91a} & $N^{2}$ \\
 \algo{MMM2D} & Arnold \& Holm \cite{arnold01b} & $N^{5/3}$ \\
 \algo{EW3DC} & Yeh \& Berkowitz \cite{yeh99a} & $N^{3/2}$ \\
 \algo{EW3DLC}& Ewald \cite{ewald21a} + \algo{ELC}& $N^{3/2}$ \\
 \algo{P3MLC} & Hockney \& Eastwood\cite{hockney88a} + \algo{ELC}& $N\log N$ \\
 \algo{MMMLC} & Sperb \& Strebel\cite{sperb01a}  + \algo{ELC}& $N\log N$ \\
 \algo{FMMLC} & Greengard \& Rhoklin \cite{greengard87a} + \algo{ELC}& $N$
\end{tabular}
\label{tab1}
\end{table}

This paper is organized as follows. In Sec. \ref{elc} the equations required
for implementation of \algo{EW3DLC} are collected and discussed. In Sec.
\ref{error} the standard Ewald error formulas are extended for anisotropic
systems and the error formula for the new layer correction term is described.
The methodology is demonstrated on several test-systems in Sec. \ref{demo} and
the efficiency compared with other 2D methods is shown in Sec. \ref{time}.
Finally the results of this paper and the outlook for computational studies of
slab geometries are summarized in Sec. \ref{disc}. In the Appendix we give
explicit formulas for the \algo{ELC}-term contributions to the force and the
diagonal part of the stress tensor.


\section{\algo{EW3DLC} in a Nutshell}

\label{elc}

The electrostatic energy of one section of a periodic slab follows from the
classical formula for the Coulomb pair potential. Omitting the prefactor ($%
1/4\pi \epsilon _{0}$), 
\begin{equation}
E=\frac{1}{2}\sum_{{\bf n}_{\parallel }}^{\prime }\sum_{i,j=1}^{N}\frac{%
q_{i}q_{j}}{|{\bf r}_{ij}+{\bf n}_{\parallel }|}.  \label{Energy}
\end{equation}
This serves as a useful point of introduction to our notation: $N$
particles with charges $q_{i}$ and positions ${\bf r}_{i}$, reside in
a simulation box of edges $L_{x}\times L_{y}\times L_{z}$. The image
boxes are denoted using the vector ${\bf n}_{\parallel }\in ({\Bbb
Z}L_{x},{\Bbb Z}L_{y},0)$ and ${\bf r}_{ij}={\bf r}_{i}-{\bf
r}_{j}$. The prime on the inner summation indicates the omission of
the primary box, ${\bf n}_{\parallel }=(0,0,0)$, when $i=j$ (the
singular case). Strictly speaking the sum over ${\bf n}_{\parallel}$
is ill-defined without specification of the order of summation. This
is a consequence of the fact that summations of the type $|{\bf
n}+c|^{-\xi }$ are conditionally convergent for $0<\xi <d$, where $d$
is the dimensional periodicity. The summation limit of relevance is an
expanding cylinder or disc such that the summation terms appear in
order of increasing $|{\bf n}_{\parallel }|$.

We convert this into a three-dimensional problem by nesting the $L_{x}\times
L_{y}\times h$\ particle volume within a larger $L_{x}\times L_{y}\times
L_{z}$\ box, thereby creating a screen for the ghost charges (the new image
charges) as described in the introduction. The total electrostatic energy $E$
is, via combination of the celebrated Ewald identity and the recent
convergence-parameter layer correction, rewritten in the computationally
convenient form: 
\begin{equation}
E=E^{(r)}+E^{(k)}+E^{(s)}+E^{(d)}+E^{(lc)},  \label{Eparts}
\end{equation}
where the individual terms are as follows,

\begin{eqnarray}
E^{(r)} &=&\frac{1}{2}\sum_{i,j=1}^{N}\sum_{{\bf n}}^{\prime }q_{i}q_{j}%
\frac{\mbox{erfc}(\alpha |{\bf r}_{ij}+{\bf n}|)}{|{\bf r}_{ij}+{\bf n}|}
\label{Real_term} \\
E^{(k)} &=&\frac{1}{2}\frac{1}{L_{x}L_{y}L_{z}}\sum_{{\bf k}\neq 0}\frac{%
4\pi }{k^{2}}e^{-k^{2}/4\alpha ^{2}}\left| \sum_{j=1}^{N}q_{j}e^{i{\bf k}%
\cdot {\bf r}_{j}}\right| ^{2}  \label{Fourier_term} \\
E^{(s)} &=&-\frac{\alpha }{\sqrt{\pi }}\sum_{i}q_{i}^{2}  \label{Self_energy}
\\
E^{(d)} &=&\frac{2\pi }{L_{x}L_{y}L_{z}}\Bigl(\sum_{i}q_{i}z_{i}\Bigr)^{2}
\label{Dipole_term} \\
E^{(lc)} &=&-\frac{8\pi }{L_{x}L_{y}}\sum_{k_{x},k_{y}>0}\frac{1}{k_{\Vert
}(e^{k_{\Vert }L_{z}}-1)}\sum_{p,q,r=1}^{2}(-1)^{p}\chi _{pqr}^{2}  \nonumber
\\
&&-\frac{4\pi }{L_{x}L_{y}}\sum_{k_{x}>0}\frac{1}{k_{x}(e^{k_{x}L_{z}}-1)}%
\sum_{p,q=1}^{2}(-1)^{p}\chi _{pq0}^{2}  \nonumber \\
&&-\frac{4\pi }{L_{x}L_{y}}\sum_{k_{x}>0}\frac{1}{k_{y}(e^{k_{y}L_{z}}-1)}%
\sum_{p,q=1}^{2}(-1)^{p}\chi _{p0q}^{2}  \label{Layer_correction}
\end{eqnarray}
where the summations run over ${\bf n}\in ({\Bbb Z}L_{x}, {\Bbb
Z}L_{y},{\Bbb Z}L_{z})$ and ${\bf k}\in 2\pi ({\Bbb Z}/L_{x},{\Bbb
Z}/L_{y},{\Bbb Z}/L_{z})$ and are order-independent. As usual, the
symbol $\alpha$ denotes the Ewald parameter which: (1) determines the
radius of the Gaussian charge distributions, (2) has units of inverse
length and (3) can take any value without changing the net result for
the energy, but has an optimum with respect to convergence of the
summations. The first three terms (Eqs. \ref{Real_term}--
\ref{Self_energy}) correspond to the usual Ewald real (r), Fourier (k)
and self (s) contributions to the energy. The dipole term $E^{(d)}$
arises from the cylindrical or slab-wise summation limit as discussed
in paper I\cite{arnold01d} and at length in the introduction. It
differs from the spherical version by a factor of $3$ and by the
replacement of the total dipole moment with its $z$-component. Again,
this dipole term applies only to vacuum boundary conditions. The final
term $E^{(lc)}$ is the energy subtraction due to all image
layers. This term was derived in paper I and is shown here in an
expanded $O(N)$\ form. These summations run over the variables
$k_{x},k_{y}\in 2\pi {\Bbb Z}/L_{x},2\pi {\Bbb Z}/L_{y}$ and the
parallel combination of these is
$k_{\Vert}=\sqrt{k_{x}^{2}+k_{y}^{2}}$.  The component abbreviations
are as follows,
\begin{equation}
\chi _{1/2,1/2,1/2}=\sum_{i=1}^{N}q_{i}\sinh /\cosh (k_{\Vert }z_{i})\sin
/\cos (k_{x}x_{i})\sin /\cos (k_{y}y_{i})
\end{equation}
and a subscript of zero stipulates omission of the corresponding
trigonometric function. For example, $\chi _{2,0,1}=\sum q_{i}\cosh
(k_{\Vert }z_{i})\sin (k_{y}y_{i})$. The corresponding equations for the
forces can be obtained by differentiation of Eqs. \ref{Real_term}--\ref
{Layer_correction}, i.e. ${\bf f}_{i}=-\nabla _{i}E$.

The derivation of the layer correction energy is a crucial part of the
method so we shall recapitulate it here in brief. The objective is to obtain
a rapidly convergent linear expression for the total energy of interaction
of the particles in the primary box with the ghost charges: 
\begin{equation}
E^{(lc)}=-\frac{1}{2}\sum_{n_{z}\neq 0}\sum_{{\bf n}_{\parallel
}}\sum_{i,j=1}^{N}\frac{q_{i}q_{j}}{|{\bf r}_{ij}+{\bf n}_{\parallel }|}.
\end{equation}
It is quite similar in appearance to the original formula (Eq. \ref{Energy}%
), and can be treated analogously by convergence factors as in the study by
Arnold and Holm \cite{arnold01b}.

We have not here written explicitly the summation order but nevertheless
this is an important consideration. \cite{sum-order} With the Poisson
summation formula applied along each periodic direction a transformation of
the in-plane variables, $n_{x}$ and $n_{y}$, into Fourier conjugates $k_{x}$
and $k_{y}$ is achieved. The derivation depends heavily on the requirement
that the system is overall neutral and results finally in

\begin{eqnarray}
E^{(lc)}=-\sum_{i,j=1}^{N}q_{i}q_{j}\Biggl(\; &&\frac{8\pi }{L_{x}L_{y}}%
\sum_{k_{x},k_{y}>0}\frac{\cosh (k_{\parallel }z_{ij})}{k_{\parallel
}(e^{k_{\parallel }L_{z}}-1)}\cos (k_{x}x_{ij})\cos (k_{y}y_{ij})+
\label{mmmsum} \\
&&\frac{4\pi }{L_xL_y}\sum_{k_{x}>0}\frac{\cosh (k_{x}z_{ij})\cos
(k_{x}x_{ij})}{k_{x}(e^{k_{x}L_{z}}-1)}+  \nonumber \\
&&\frac{4\pi }{L_yL_x}\sum_{k_{y}>0}\frac{\cosh (k_{y}z_{ij})\cos
(k_{y}y_{ij})}{k_{y}(e^{k_{y}L_{z}}-1)}\Biggr).  \nonumber
\end{eqnarray}
It is fortunate (in contrast to \algo{EW2D}) that the coordinates appear as
they do because after decomposition of the $\cos$ and $\cosh$ terms, the
$N^{2}$ summation collapses, yielding the order $N$ Eq.
\ref{Layer_correction}. For reference purposes we have assembled in the
appendix the full expressions for the force and the diagonal parts of the
stress tensor due to the \algo{ELC}-term.


\section{Error Estimates}

\label{error}

The most important measures of error in an electrostatic calculation are the
root mean squared error in the particle forces,

\begin{equation}
\Delta f:=\Biggl(\frac{1}{N}\sum_{i=1}^{N}\Delta {\bf f}_{i}^{2}\Biggr)%
^{1/2},  \label{force_error}
\end{equation}
where $\Delta {\bf f}_{i}$ is the difference between the computed and exact
force on particle $i$, and the absolute value of the error in the total
electrostatic energy, $\Delta E$. We will discuss mainly the former, which
is relevant to molecular dynamics simulations.

The total error can be separated into terms originating from the cutoffs
used for the Ewald and layer correction,

\begin{equation}
\Delta f\cong \sqrt{\Delta f_{r}^{2}+\Delta f_{k}^{2}+\Delta f_{l}^{2}},
\label{err_contrib}
\end{equation}
where the individual components are determined in the same way as in Eq. \ref
{force_error}. Analytical error estimates for the real- and reciprocal-space
terms for a cubic simulation box have been previously derived.\cite
{kolafa92a,petersen95a} Shortly we will extend these results to obtain the
formulas for arbitrary box sizes, but we write first the result,

\begin{eqnarray}
\Delta f_{r}(\alpha {,}r_{c}) &=&\frac{2}{\sqrt{L_{x}L_{y}L_{z}}}\frac{Q^{2}%
}{\sqrt{N}}\frac{\exp {[-}\alpha ^{2}r_{c}^{2}{]}}{\sqrt{r_{c}}},
\label{Real_error} \\
\Delta f_{k}(\alpha {,}k_{c}) &=&\frac{2\alpha}{\sqrt{\pi L_{y}L_{z}}}\frac{%
Q^{2}}{\sqrt{N}}\frac{\exp {[-(}\pi k_{c}/\alpha L_{x}{)^{2}]}}{\sqrt{k_{c}}}%
,  \label{Fourier_error}
\end{eqnarray}
where $Q^{2}=\sum_{i=1,N}q_{i}^{2}$, $r_{c}$\ is the real-space minimum
image cutoff (typically but not necessarily $\leq \frac{1}{2}\min
[L_{x},L_{y},L_{z}]$) and $k_{c}$\ is an integer defining the truncation of
the reciprocal-space summation. These naturally reduce to the standard
formulas when the box is cubic. The optimum set of convergence parameters is
obtained by setting $\Delta f_{r}^{2}=\Delta f_{k}^{2}$. This reduces the
number of independent variables to one yielding implicit relations $\alpha
(k_{c}),r_{c}(k_{c})$. Then one can easily scan $k_{c}$ to find the optimum
(i.e. fastest) $k_{c}^{\ast }$.

If the reciprocal space computation is instead handled by FFT methods such as
\algo{P3M}, inaccuracies arise due to the assignment and de-assignment of the
charges onto a mesh. The error for \algo{P3M} has been estimated (for a cubic
system) by
\begin{equation}
\Delta f_{P3M}(\alpha ,M,P)=\frac{Q^{2}}{L^{2}\sqrt{N}}\left( \frac{\alpha L%
}{M}\right) ^{P}\sqrt{\alpha L\sqrt{2\pi }\sum_{k=0}^{P-1}a_{k}^{(P)}\left( 
\frac{\alpha L}{M}\right) ^{2k}}  \label{p3m_error}
\end{equation}
where $M$ is the number of mesh points per edge and $P$ is the interpolation
order. The expansion coefficients, $a_{k}^{(P)}$ are constants given by
Deserno and Holm.\cite{deserno98b} A more accurate numerical error estimate
is also given by these authors but Eq. \ref{p3m_error} is useful since it
captures the function form of the error.

\begin{figure}[tbp]
\begin{center}
\includegraphics[scale=0.5]{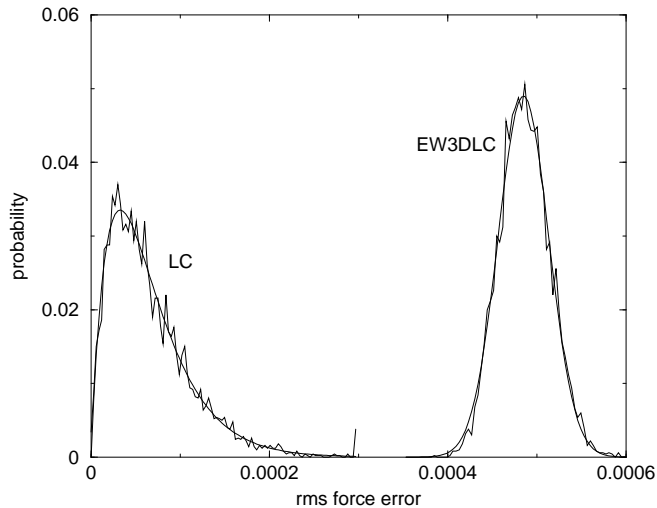}
\end{center}
\caption{
Error distributions for \algo{EW3DLC}, $p(\Delta f)$, and layer correction
, $p(\Delta f_{lc})$, for a 100 particle system of physical dimensions
 $1.0 \times 1.0 \times 0.9$. Convergence parameters for \algo{EW3DLC}:
$\alpha=9.3$, $r_c=0.36$, $k_c=10$, $L_z=2.0$, $\Delta f_{lc}<10^{-3}$;
for LC: $L_z=1.0$, $\Delta f_{lc}<10^{-2}$.
}
\label{lc_err_dist}
\end{figure}

The Ewald terms have the advantage of well-behaved error distributions. This
is not the case when turning to the layer correction term. Consider for
example the probability distribution of $\Delta f$ and $\Delta f_{lc}$ over
the ensemble of random particle configurations as shown in Fig. \ref
{lc_err_dist}. The former distribution was generated under conditions of
negligible layer correction error while the latter distribution is for layer
correction error itself under conditions where it is significant (i.e. small
gap). The distribution of the total \algo{EW3DLC} error fits very well to a
Gaussian curve. The error distribution of the layer correction term on the
other hand is skewed to the left and fits well to a stretched exponential $%
p(x)\propto x\exp (-ax)$. The extended tail of this distribution means that
exceptions to the average error will occur with non-negligible frequency.
This feature makes it difficult to derive an accurate RMS error estimate for
the layer correction. But what is possible and more useful is the derivation
of a stricter error criterion, namely an upper bound on the RMS error. A
rigorous upper bound was derived in the first paper yielding, 
\begin{eqnarray}
\Delta f_{lc}(l_{c},L_{z}) &\leq &\frac{Q^{2}}{\sqrt{N}}\frac{\sqrt{3}}{%
2(e^{2\pi l_{c}L_{z}/L}-1)}\biggl(\left( \frac{2\pi l_{c}+4}{L}+\frac{1}{%
L_{z}-h}\right) \frac{e^{2\pi l_{c}h/L}}{L_{z}-h}  \nonumber \\
&&+\left( \frac{2\pi l_{c}+4}{L}+\frac{1}{L_{z}+h}\right) \frac{e^{-2\pi
l_{c}h/L}}{L_{z}+h}\biggr),
\end{eqnarray}
where $l_{c}$ is the integer-valued radial cutoff for truncation of the
layer correction forces.

Before leaving this topic a reminder is due of the bias present in slab
geometries. Both the Ewald and layer correction terms have an error bias
with respect to the $z$-position of a particle. The error tends to be higher
near the walls of the system as if it were ``hotter'' in those regions. In
fact, one can imagine a simulation in which the overall RMS error in the
forces is smaller than the RMS random force but due to this bias, the RMS
force-error at the surfaces is larger than the thermal noise, amounting to
true hot zones at the surfaces. Therefore it is recommended to exercise some
caution in the treatment of these effects. The layer correction error in
particular has a strong surface bias, and it is therefore quite appropriate
to use an upper bound in place of an RMS error estimate. In other words, the
layer correction should be carried out to higher accuracy than the real and
reciprocal space parts.

\subsection*{Anisotropic Real-Space Error}

Let us write down the real-space force first, 
\begin{equation}
{\bf f}_{i}^{(r)}=\sum_{j=1}^{N}q_{i}q_{j}\sum_{{\bf n}}^{\prime }{\bf r}%
_{ijn}\left[ \frac{2\alpha }{\sqrt{\pi }}\frac{e^{-\alpha ^{2}r_{ijn}^{2}}}{%
r_{ijn}^{2}}+\frac{\mbox{erfc}\left( \alpha |{\bf r}_{ijn}|\right) }{|{\bf r}%
_{ijn}|r_{ijn}^{2}}\right].  \label{real_space_force}
\end{equation}
This is always summed using a minimum image spherical cutoff
$|r_{mjn}|<r_{c}$ regardless of the system geometry. Therefore the
error in truncation of Eq. \ref{real_space_force} for the force on
particle $i$ is
\begin{equation}
\delta {\bf f}_{i}=q_{i}\sum_{j=1}^{N}\sum_{{\bf n}:|{\bf r}%
_{ijn}|>r_{c}}q_{j}{\bf r}_{ijn}g(|{\bf r}_{ijn}|),
\end{equation}
where $g$ is an abbreviation for the term in square brackets. We square this
and obtain 
\begin{equation}
(\delta {\bf f}_{i})^{2}=q_{i}^{2}\sum_{j\neq k}q_{j}q_{k}\sum_{|{\bf r}%
_{1}|>c}\sum_{|{\bf r}_{2}|>r_{c}}{\bf r}_{1}\cdot {\bf r}_{2}g(|{\bf r}%
_{1}|)g(|{\bf r}_{2}|)+q_{i}^{2}\sum_{j=1}^{N}q_{j}^{2}\sum_{{\bf n}:|{\bf r}%
_{1}|>r_{c}}^{2}{\bf r}_{1}^{2}g^{2}(|{\bf r}_{1}|),
\end{equation}
where ${\bf r}_{1},{\bf r}_{2}$ are the vectors pointing from particle $i$
to the $n^{th}$ images of $j$ and $k$ respectively. Since $g$ is an even
function and $q_{1},q_{2}$ will take negative and positive values the
average of this quantity over all the $i$ particles in the system contains,
in the limit of large, uncorrelated $N$, only the diagonal terms. Thus 
\begin{equation}
\left\langle (\delta {\bf f}_{i})^{2}\right\rangle
=q_{i}^{2}\sum_{j=1}^{N}q_{j}^{2}\sum_{n:|r_{1}|>r_{c}}{\bf r}_{1}^{2}g^{2}(|%
{\bf r}_{1}|).
\end{equation}
Moreover in the statistical limit the summation $\Sigma _{n}$ is independent
of particles $i$ and $j$ and we have 
\begin{equation}
\left\langle (\delta {\bf f}_{i})^{2}\right\rangle ^{1/2}=|q_{i}|Q\Sigma
_{n}^{1/2},
\end{equation}
where $Q^{2}=\sum q_{j}^{2}$. We will proceed by treating the summation as a
spherical integral. First we write the variable explicitly 
\begin{equation}
|{\bf r}_{1}|=\sqrt{(x+sL_{x})^{2}+(y+tL_{y})^{2}+(z+uL_{z})^{2}}=\rho
\end{equation}
where $s,t,u\in {\Bbb Z}$. In Cartesian coordinates we have the following
integration 
\begin{equation}
\Sigma _{n}\approx \frac{8}{L_{x}L_{y}L_{z}}\int_{\widetilde{c_{s}}}^{\infty
}d\widetilde{s}\int_{\widetilde{c_{t}}}^{\infty }d\widetilde{t}\int_{%
\widetilde{c_{u}}}^{\infty }d\widetilde{u}\rho ^{2}g^{2}(\rho )=\frac{4\pi }{%
LL_{y}L_{z}}\int_{r_{c}}^{\infty }d\rho \rho ^{4}g^{2}(\rho )
\end{equation}
where we have applied $\widetilde{s}=s/L_{x}$ etc.

The final step involves the application of the asymptotic integral
expansion, 
\begin{equation}
\int_{z}^{\infty }f(x)e^{-ax^{2}}dx\approx \frac{f(z)e^{-az^{2}}}{2az}
\end{equation}
where the validity of this relation is discussed by Kolafa \& Perram. We can
apply this formula twice to obtain our mean-squared particle error; 
\begin{equation}
\mbox{erfc}(\alpha \rho )\approx \frac{e^{-\alpha ^{2}\rho ^{2}}}{2\alpha
\rho }
\end{equation}
and 
\begin{equation}
\int_{c}^{\infty }d\rho e^{-2\alpha ^{2}\rho ^{2}}\left[ \frac{2\alpha }{%
\sqrt{\pi }}+\frac{1}{2\alpha \rho ^{2}}\right] ^{2}\approx \left( \frac{1}{%
\sqrt{\pi r_{c}}}+\frac{1}{4\alpha ^{2}r_{c}^{5/2}}\right) ^{2}e^{-2\alpha
^{2}r_{c}^{2}}.
\end{equation}
Finally we can obtain the root mean squared force error, 
\begin{equation}
\Delta f=\sqrt{N^{-1}\sum_{m=1}^{N}\delta f_{m}}\approx \frac{Q}{\sqrt{N}}%
\left( \frac{\left\langle (\delta f_{m})^{2}\right\rangle }{q_{m}^{2}}%
\right) ^{1/2}=\frac{2Q^{2}}{\sqrt{N}}\frac{1}{\sqrt{L_{x}L_{y}L_{z}}}\left( 
\frac{1}{\sqrt{r_{c}}}+\frac{\sqrt{\pi }}{4\alpha ^{2}r_{c}^{5/2}}\right)
e^{-\alpha ^{2}r_{c}^{2}}.
\end{equation}
Neglecting the smaller $\alpha ^{-2}$ term yields Eq. \ref{Real_error}.

\subsection*{Anisotropic Reciprocal-Space Error}

First we write the generalized anisotropic $k$-space force on particle $i$, 
\begin{equation}
{\bf f}_{i}^{(k)}=\frac{q_{i}}{L_{x}L_{y}L_{z}}\sum_{j}q_{j}\sum_{{\bf k}%
\neq 0}\frac{4\pi {\bf k}}{k^{2}}\exp \left( -\frac{k^{2}}{4\alpha ^{2}}%
\right) \sin ({\bf k}\cdot {\bf r}_{ij}),  \label{kforce}
\end{equation}
where again ${\bf k}\in 2\pi ({\Bbb {Z}}/L_{x},{\Bbb {Z}}/L_{y},{\Bbb {Z}}%
/L_{z}{\Bbb )}$.

The summation will be approximated using a cutoff $\tau$ and therefore only
terms satisfying 
\begin{equation}
{\bf k}^2\le \tau^2
\end{equation}
will be added. This equation defines the interior of an ellipsoid whose axes
are $L_x/2\pi$, etc.

Now we compose the cutoff error for a single pair of particles, for
which it will prove useful to write using complex notation. Using the
fact the function in front of the $\sin$ is odd and that half of the
vectors ${\bf k}$ are identical but opposite in sign to the other half
we obtain
\begin{equation}
\delta {\bf f}_{ij}=-\frac{q_{i}q_{j}}{L_{x}L_{y}L_{z}}\sum_{|{\bf k}|>\tau }%
\frac{4\pi {\bf k}}{k^{2}}\exp \left( -\frac{k^{2}}{4\alpha ^{2}}\right)
i\exp (i{\bf k}\cdot {\bf r}_{ij}).
\end{equation}
\bigskip

Taking the square of this yields 
\begin{equation}
(\delta {\bf f}_{ij})^{2}=-\left( \frac{q_{i}q_{j}}{L_{x}L_{y}L_{z}}\right)
^{2}\sum_{|{\bf k}_{1}|>\tau }\sum_{|{\bf k}_{2}|>\tau }\frac{16\pi ^{2}{\bf %
k}_{1}\cdot {\bf k}_{2}}{k_{1}^{2}k_{2}^{2}}e^{-(k_{1}^{2}+k_{2}^{2})/4%
\alpha ^{2}}\exp (i{\bf k}_{1}\cdot {\bf r}_{ij})\exp (i{\bf k}_{2}\cdot 
{\bf r}_{ij}).
\end{equation}
At this point we remind ourselves that, in spite of how it looks, this
expression is indeed real and positive-valued. We want to obtain the average
of this function over all the particle pairs in the system. In the limit of
many uncorrelated particles the following expression is exact, 
\begin{equation}
\left\langle \exp (i{\bf k}_{1}\cdot {\bf r}_{ij})\exp (i{\bf k}_{2}\cdot 
{\bf r}_{ij})\right\rangle _{ij}=\delta ({\bf k}_{1},-{\bf k}_{2}).
\end{equation}
The details of this require the observation that $%
<s_{1}c_{2}>=<s_{1}s_{2}>=<c_{1}c_{2}>=0$ for ${\bf k}_{1}\neq {\bf
k}_{2},\neq -{\bf k}_{2}$ and $<s^{2}>=<c^{2}>$ where the
abbreviations $s_{1}=\sin ({\bf k}_{1}\cdot {\bf r}_{ij})$ and
$c_{1}=\cos ({\bf k}_{1}\cdot {\bf r}_{ij})$ have been used. Finally
we reach a greatly simplified version of the root-mean-squared
pair-wise error:
\begin{equation}
\left\langle (\delta {\bf f}_{ij})^{2}\right\rangle ^{1/2}=\frac{%
|q_{i}||q_{j}|}{L_{x}L_{y}L_{z}}\left( \sum_{|{\bf k}|>\tau }\frac{16\pi ^{2}%
}{k^{2}}e^{-k^{2}/2\alpha }\right) ^{1/2}.
\end{equation}

Now we will be somewhat explicit in treating the summation above. We rewrite
the summation in terms of its fundamental variables and length scales, 
\begin{equation}
\sum_{|{\bf k}|>\tau }\frac{e^{-k^{2}/2\alpha ^{2}}}{k^{2}}=\frac{L_{x}^{2}}{%
4\pi ^{2}}\sum_{\rho ^{2}>k_{c}^{2}}\frac{\exp [-a\rho ^{2}]}{\rho ^{2}}.
\end{equation}
Here we are taking $a=2\pi ^{2}/\widetilde{\alpha }^{2}=2\pi ^{2}/(\alpha
L)^{2}$, $\lambda =L_{x}/L_{z}$, $\gamma =L_{x}/L_{y}$, $\rho
^{2}=l^{2}+m^{2}\gamma ^{2}+n^{2}\lambda ^{2}$, $l,m,n\in {\Bbb Z}$ and $%
k_{c}$ is the integer cutoff which defines an ellipsoidal solid upon ${\Bbb Z%
}^{3}$. With this notation, all of the units of the expression are carried
by the factor $L_{x}^{2}$. This equation can be approximated by integration;
in Cartesian coordinates 
\begin{equation}
\approx 8\frac{L_{x}^{2}}{4\pi ^{2}}\int_{S1}dl\int_{S2}dm\int_{S3}dn\frac{%
\exp [-a(l^{2}+m^{2}\gamma ^{2}+n^{2}\lambda ^{2})]}{l^{2}+m^{2}\gamma
^{2}+n^{2}\lambda ^{2}},
\end{equation}
where the limits which we do not write here, will collapse into a much
simpler form.
Conversion to spherical coordinates requires only the substitutions $%
\widetilde{n}=\lambda n$, $\widetilde{m}=\gamma m$ and we have 
\begin{equation}
\frac{L_{x}^{2}}{\pi }\gamma ^{-1}\lambda ^{-1}\int_{k_{c}}^{\infty }\rho
^{2}d\rho \frac{\exp (-a\rho ^{2})}{\rho ^{2}}\approx \frac{L_{x}^{2}}{\pi }%
\gamma ^{-1}\lambda ^{-1}\frac{\exp (-ak_{c}^{2})}{2ak_{c}}.
\end{equation}
To summarize the results of this section, we have found 
\begin{equation}
\sum_{|{\bf k}|>\tau }\frac{e^{-k^{2}/2\alpha ^{2}}}{k^{2}}\approx L^{2}%
\frac{\widetilde{\alpha }^{2}}{4\pi ^{3}\gamma \lambda l_{c}}\exp (-2\pi
^{2}k_{c}^{2}/\widetilde{\alpha }^{2}),
\end{equation}
a result that was obtained by imposing a single-parameter ellipsoidal cutoff
which makes simple spherical integration possible. That is to say, if we had
not restricted the degrees of freedom of the cutoff object, then we would be
faced with the daunting task of ellipsoidal integration. Two assumptions
were made, namely that the function varies slowly enough across the grid
points so that integration is valid, and secondly that the integral theorem
mentioned by Kolafa \& Perram is accurate.

We can write the RMS error, first we have the pair-wise error approximated
as 
\begin{equation}
\left\langle (\delta {\bf f}_{ij})^{2}\right\rangle ^{1/2}=\frac{%
|q_{i}||q_{j}|}{L_{y}L_{z}}\frac{2}{\sqrt{\pi }}\frac{\widetilde{\alpha }}{%
\sqrt{\gamma \lambda k_{c}}}\exp (-\pi ^{2}k_{c}^{2}/\widetilde{\alpha }%
^{2});
\end{equation}
which leads to 
\begin{equation}
\Delta f\approx \sqrt{\frac{1}{N}\sum_{i=1}^{N}\sum_{j\neq
i}^{{}}\left\langle (\delta {\bf f}_{ij})^{2}\right\rangle }\approx \frac{%
Q^{2}}{\sqrt{N}}\frac{2}{\sqrt{\pi }}\frac{1}{L_{y}L_{z}}\frac{\widetilde{%
\alpha }}{\sqrt{\gamma \lambda k_{c}}}\exp (-\pi ^{2}k_{c}^{2}/\widetilde{%
\alpha }^{2}).
\end{equation}


\section{Demonstration of Accuracy}

\label{demo}

In this section we demonstrate the credibility of the \algo{EW3DLC} technique
by means of several standardized test cases that have been used in other
literature. These fall into two categories: (1) two-particle systems and (2) a
single charge elevated above a planar array of 25 charges. These cases are
known to emphasize the possible errors arising from the non-periodicity in the
$z$-direction. Secondly, several tests of the error formulas are carried out
to see that they are indeed applicable to non-cubic, inhomogeneous systems.
For these tests it is sufficient to use a standard three-dimensional Ewald
summation.

\begin{figure}[tbp]
\begin{center}
\includegraphics[scale=0.5]{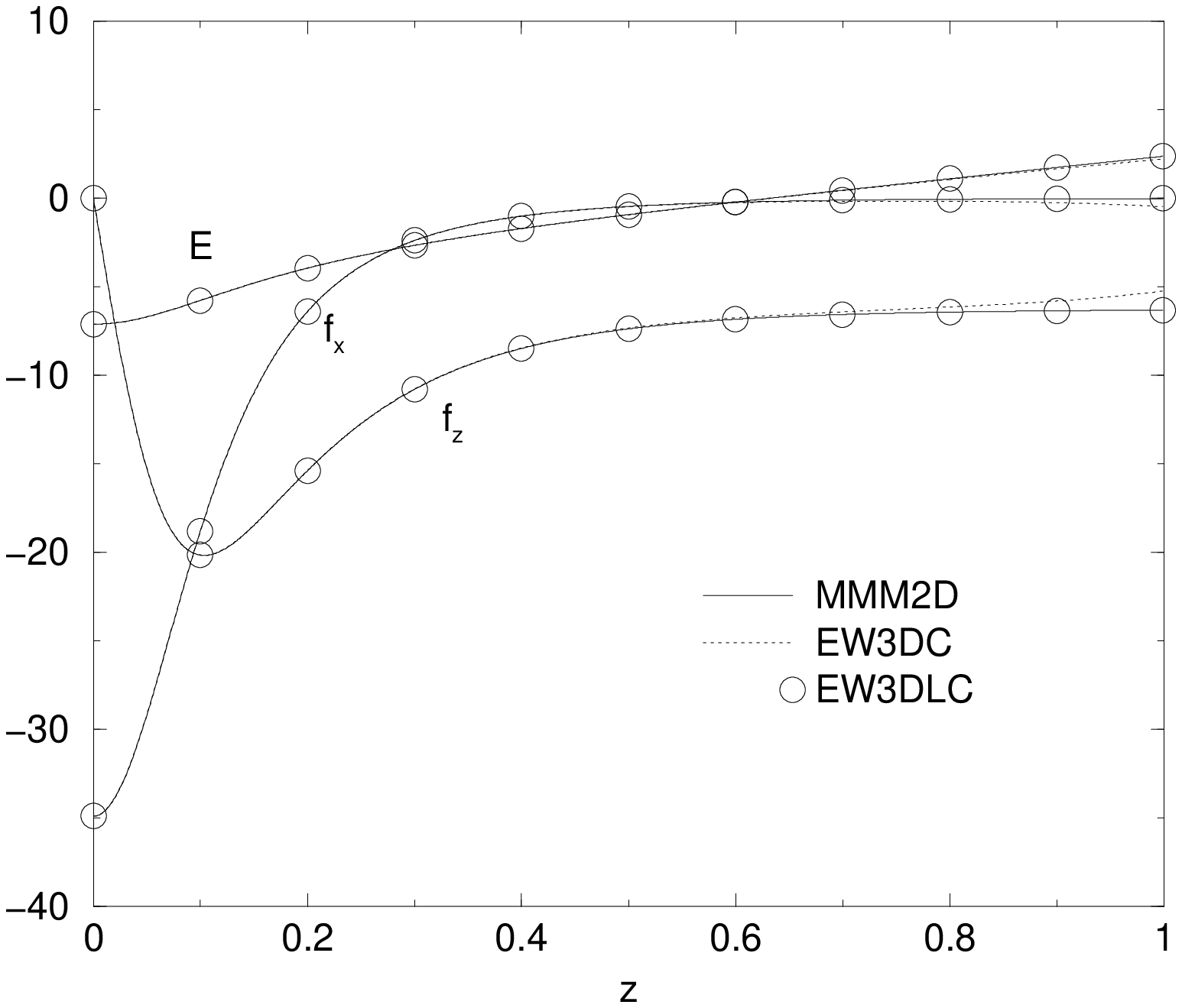}
\includegraphics[scale=0.5]{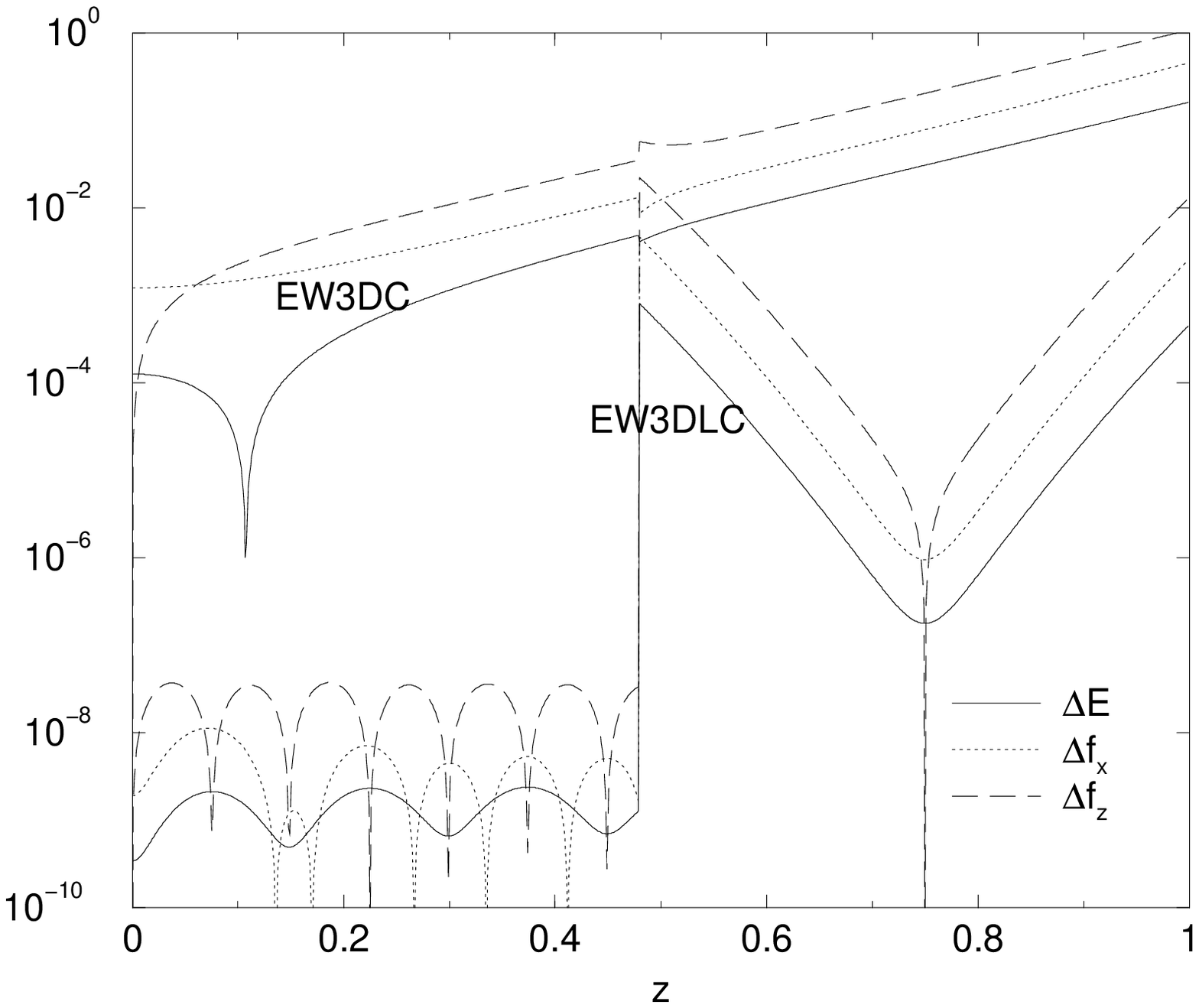}
\end{center}
\caption{Energy and force in a two-particle system: ${\bf r}_{1}=(0,0,0)$, $
{\bf r}_{2}=(0.1,0.1,z)$, $q_{1}=-1$ and $q_{2}=1$. As a reference, a
well-converged ($\Delta f < 10^{-10}$) \algo{MMM2D} calculation is used.
Convergence parameters:
$\alpha =5.0$, $k_{c}=10$, $r_{c}=0.5$, $L_{z}=1.5$, $\Delta f_{lc}<10^{-10}$.
(a) Energy and forces. (b) Absolute error in energy and forces. }
\label{twoparticles}
\end{figure}

In Fig. \ref{twoparticles} the results from a two-particle system are given.
In this system the first particle is fixed and the second particle is moving
in the $z$ direction. For the remainder of this paper we will use $%
L_{x}=L_{y}=L$ and unless otherwise noted $L$ is set equal to unity. In
principle there is no limitation on the accuracy of the \algo{EW3DLC} or
\algo{EW3DC} methods. The only visible errors (Fig. \ref{twoparticles}a) occur
for the \algo{EW3DC} method as $z\rightarrow 1$ since the particle begins to
approach the nearest image layer. This error can be eliminated (at some
computational cost) by increasing $L_{z}$, but we are interested in
illustrating the point.

A richer picture is obtained by plotting the errors on a semi-log scale (Fig.
\ref{twoparticles}b). The \algo{EW3DLC} curves are complex but these quirks
vanish already in 3 or 4 particle systems and are more a signature of the
underlying features of the Ewald equations. The key to the explanation of
these features is the spherical minimum image cutoff which we chose as
$r_{c}=0.5$. When $z$ reaches $0.48$, the particles are no longer in view of
each other, nor are any of the images in view. This leads on the right side of
the plot to a significant underestimate of the real-space energy and force and
is dominated by the exponential in Eq. \ref{Real_term}. The left half of the
\algo{EW3DLC} curves are dominated by Fourier-space error which is
trigonometric with a period of $(k_{c}+1)L/L_{z}$ due to the complex
exponential in Eq. \ref{Fourier_term}. Thus the final conclusion is that the
basic two-particle curves exhibit nothing problematic in terms of the force
and energy signatures.

The next test-system is the 26-particle one originally used by Widmann and
Adolf \cite{widmann97a}: a square array of 25 particles in the $z=0$ plane
with coordinates $(x,y,0)$ where $x,y\in (0.1,0.3,0.5,0.7,0.9)$, and the
26th particle hovers above the center at $(0.5,0.5,0.2)$. 
The values reported previously are easily reproduced using a layer 
correction. For example, setting 
$(\alpha ,r_{c},k_{c}, L_z)=(15.0,0.40,15,0.80)$ produces 
$-86.5655$ and $-10.3642$ for the total energy and the $z$-component 
force of the 26th particle respectively. For the layer correction 
the the maximal pairwise error was set at $10^{-6}$. The layer correction 
contribution in this case is important even though the box edge
is four times the layer thickness,
amounting to $-0.178234$ and $0.476241$ for the force 
and energy respectively.

With these basic properties of the proposed algorithm established we may
begin to probe its applicability to realistic systems. To this end we
generate a ``random'' system of 100 unit charges (overall electrically
neutral). We have followed the same convention as Ref. \cite{deserno98a} so
that these systems can be reproduced by other researchers. A set of 300
random numbers $n_{1},n_{2},\dots ,n_{300}$ between 0 and 1 and the
coordinates of the charges are taken as ${\bf r}%
_{1}=(n_{1}L,n_{2}L,n_{3}L_{z})$, ${\bf r}_{2}=(n_{4}L,n_{5}L,n_{6}L_{z})$,
... The system was checked to make sure that there are no problematically
small spacings. We use this system to check the applicability of the
anisotropic error formulas for varying shape and inhomogeneity and in the
next section to compare the speed of various algorithms.

\begin{figure}[tbp]
\begin{center}
\includegraphics[scale=0.45]{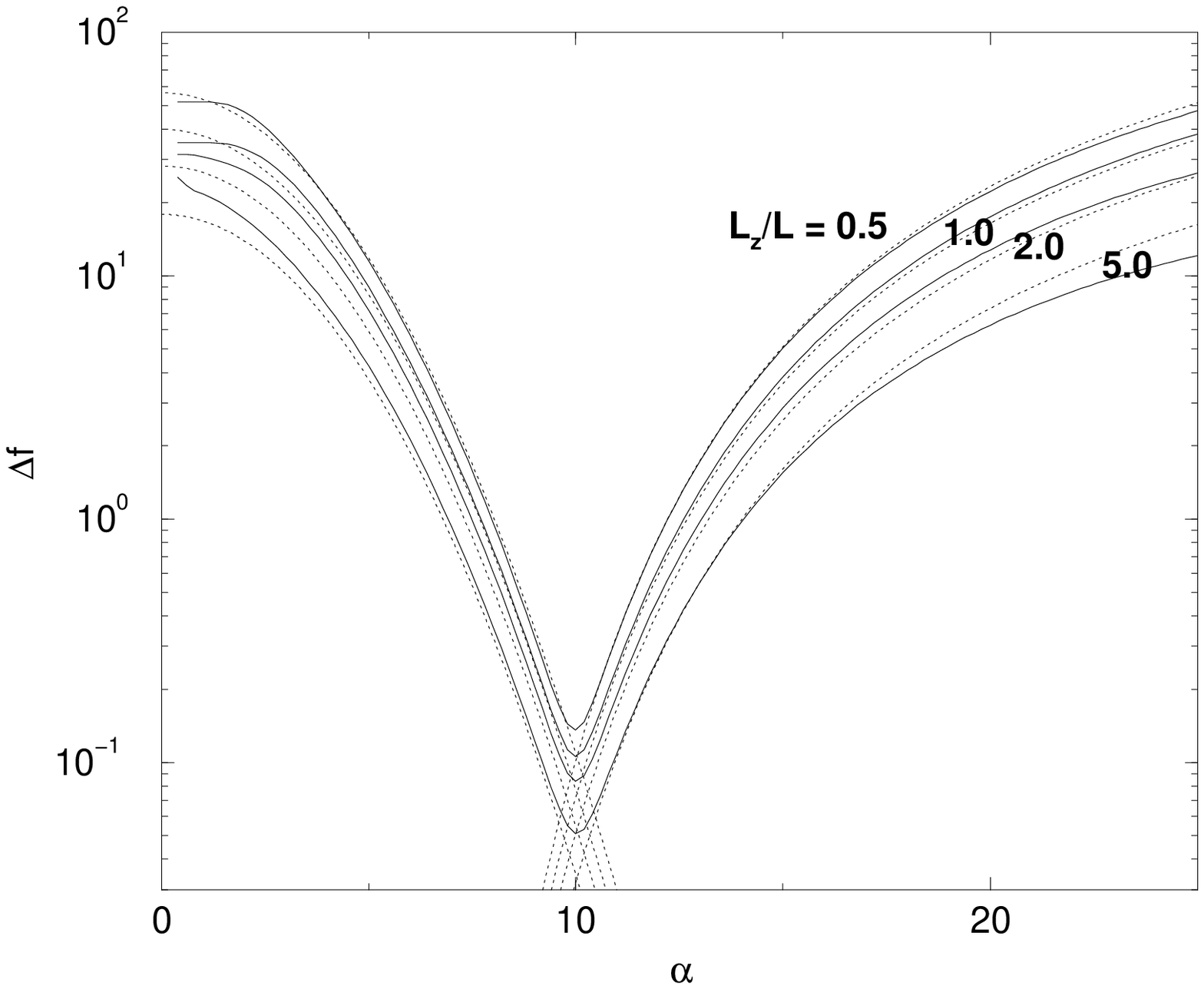}
\includegraphics[scale=0.45]{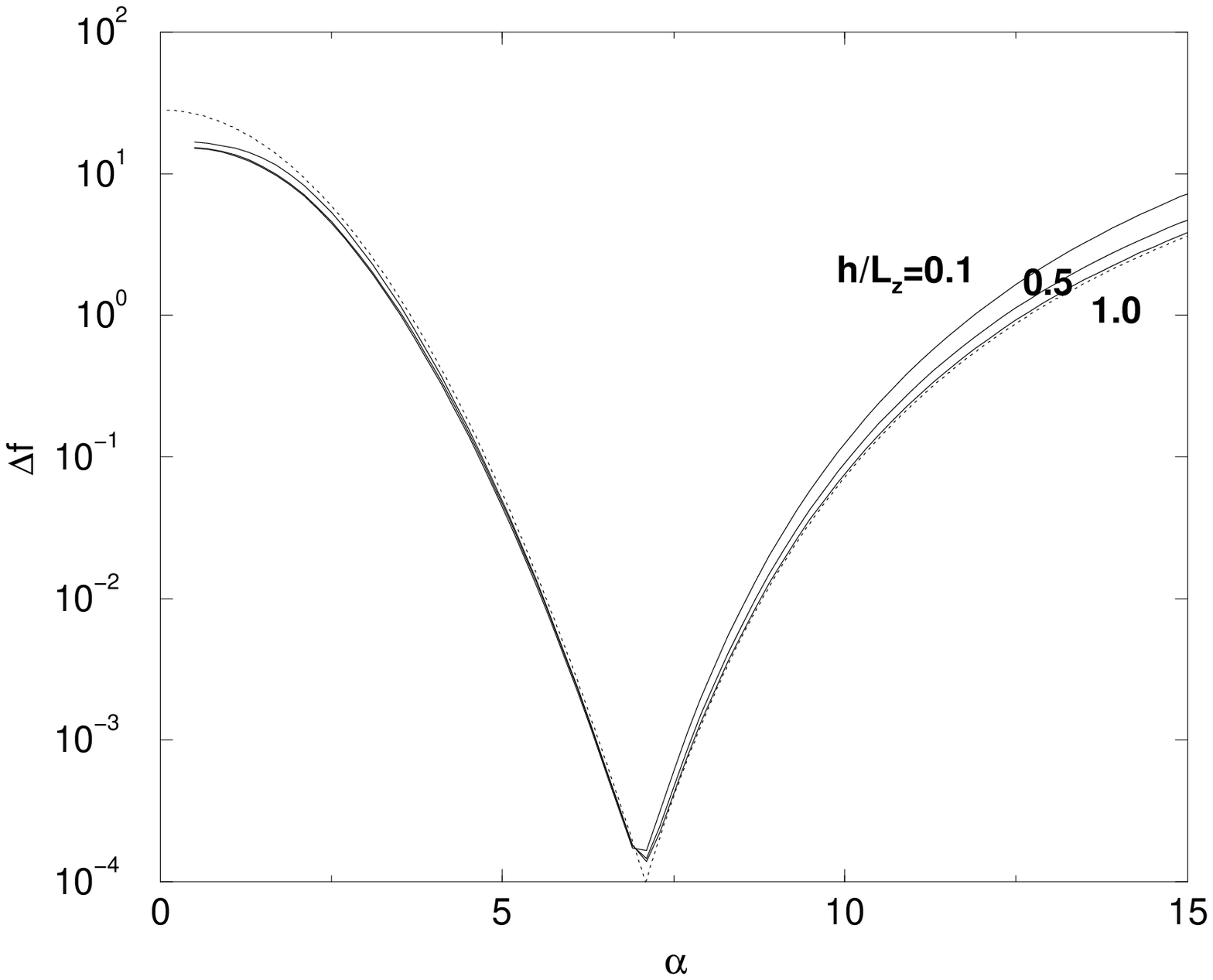}
\end{center}
\caption{Dependence of the error estimates on the box shape (a) and the
inhomogeneity (b) for standard three-dimensional Ewald summation in a
randomized 100 particle system. (a) The aspect ratio $L_z/L$ of the box is
indicated on each curve. The solid lines indicate computations $(
k_c,r_c)=(8,0.25)$ and the dotted lines are error estimates from
Eqns. \ref{Real_error}--\ref{Fourier_error}. (b) The fill ratio $h/L_z$ is
indicated on each curve. Computations (solid lines) are for $(8,0.5)$
and the dotted lines show the error estimates.}
\label{ew3d}
\end{figure}

For the error formulas it is sufficient to consider the standard Ewald
summation (\algo{EW3D}). In Fig. \ref{ew3d}a the results from a standard Ewald
summation are compared to the error formulas Eqs. \ref{Real_error}--
\ref{Fourier_error}.  For the reference forces we also used Ewald summation,
convergent to 8 digit accuracy in the total energy. The error formulas are
satisfactory for both cigar- and pancake-shaped systems (i.e. $L_{z}>1$ and
$L_{z}<1$ respectively).

The second concern is the validity of the error formulas when part of the
system is empty. This is tested for 10, 50 and 100\% filled systems in Fig. 
\ref{ew3d}b. There is no dependence on $h\ $in the error formulas and
therefore the error curves should all be the same. In fact there is good
agreement and the most substantial deviations occur in the unimportant large-%
$\alpha $ region. These deviations are evidence of only a small effect of
\algo{EW3DLC}-type inhomogeneity on the error formulas.

To summarize, we have shown that the \algo{EW3DLC} method produces accurate
and well controlled results and can be used for a broad range of shape and
inhomogeneity. This is an important advantage over the \algo{EW3DC} method
which is more restrictive on the selection of $L_{z}$ even if the error
estimate that we developed in the first paper is used to control the error.


\section{Computing Time}

\label{time}

Of course the critical consideration, once issues of accuracy are established,
is the speed with which a given algorithm can be implemented.  This makes the
\algo{P3MLC} the most efficient choice for large $N$ but for smaller systems
other algorithms such as \algo{EW3DLC} or \algo{MMM2D} may have superior speed
due to a lower overhead. This section attempts to give some quantitative
information on the break-even points between these various algorithms. As
usual, the relative times vary slightly with the programming efficiency and
the software and hardware of the computer. All times in this study were
collected on a Compaq XP1000 workstation.

To optimize the time required for such a many-parameter computation, it is
necessary to understand the scaling of time with the variables -- most
importantly with the number of charges, $N$. Here a brief, qualitative
discussion of the minimization is made. This discussion tends to be more
transparent (with the benefit of hindsight) yet less rigorous than more
detailed versions available elsewhere.

The computing time of the real-space part will in general require $0.5N\rho
(4\pi r_{c}^{3}/3)$ operations where $\rho =N/L^{2}h$ (large $h$) is the
density. Similarly the reciprocal-space calculation scales as $%
NL_{z}k_{c}^{3}$ since increasing $L_{z}$ requires a proportionate increase
of the $z$-component reciprocal space cutoff. The layer correction scales as 
$Nl_{c}^{2}$ since it is a cylindrical summation. Thus the overall time
required scales as 
\begin{equation}
t_{elc}=a_{1}N^{2}r_{c}^{3}+a_{2}NL_{z}k_{c}^{3}+a_{3}Nl_{c}^{2}
\label{time_ewald}
\end{equation}
As is evident in plots like Fig. \ref{ew3d}, the sharpness of the crossover
between the real- and reciprocal-space accuracy regimes is such that the
choice of $\alpha $, $k_{c}$ and $r_{c}$ will be near the optimum if $\Delta
f_{r}\approx \Delta f_{k}$. The layer correction cutoff can be determined
independently from the others and moreover is only a weak function of $L_{z}$
for large $(L_{z}-h)$. Eq. \ref{time_ewald} is then interpreted as a
function of two variables, for instance the reciprocal cutoff and the
desired accuracy $t_{elc}=f(k_{c},\Delta f)$. In order to neutralize the
change in accuracy in Eqs. \ref{Real_error}--\ref{Fourier_error}, the Ewald
parameter should scale inversely with the real cutoff and linearly with the
Fourier cutoff, i.e. $\alpha \sim 1/r_{c}\sim k_{c}$. Thus we can strike a
compromise or balance between the first two terms of Eq. \ref{time_ewald} by
fixing 
\begin{equation}
r_{max}\sim N^{-1/6},k_{max}\sim N^{1/6}  \label{ewald_scaling}
\end{equation}
which reduces the overall scaling of the Ewald calculation to the well-known 
$O(N^{3/2})$. Note that the inclusion of the layer correction does not
influence the scaling since it will only add a term linear in $N$ to Eq. \ref
{time_ewald}. Furthermore, it becomes a smaller part of the overall
calculation as $N$ increases.

A similar analysis applies to mesh-based algorithms, of which we will deal
with the \algo{P3M} method. The major difference is that the number of
operations required to perform a FFT is independent of the number of charges
and only depends on the number of mesh points as $M\log M$ per direction in
space.  The other part of the mesh routine is the charge assignment via
polynomial interpolation. The order of interpolation, $P$, represents the
number of grid points in each direction that a given charge is distributed
onto.  Therefore it is much like a cutoff parameter and the number of
operations per charge scales as $(P+1)^{3}$. The time required for a
\algo{P3M} computation is
\begin{equation}
t_{p3mlc}=a_{1}N^{2}r_{c}^{3}+a_{4}L_{z}(M\log
M)^{3}+a_{5}N(P+1)^{3}+a_{3}Nl_{c}^{2}  \label{p3m_time}
\end{equation}

An intuitive analysis of Eq. \ref{p3m_time} is more difficult but analogous
to the preceding reasoning (and benefits from foreknowledge of the answer).
Again we attempt to balance the $N$-exponent in the first two terms but this
time a substantially better result is possible due to the much smaller
exponent in the second term ($0$). From Eq. \ref{p3m_error}, the choice of $%
\alpha \sim M$ will roughly neutralize the effect of $M$ and thus the link
between the mesh size and real cutoff is $r_{c}\sim 1/M$. The best choice
for the scaling of these terms is 
\begin{equation}
r_{c}\sim \frac{(\log N)^{1/2}}{N^{1/3}},M\sim \frac{N^{1/3}}{(\log N)^{1/2}}
\label{p3m_scaling}
\end{equation}
resulting in an overall scaling for the time of $O(N(\log {N})^{3/2})$.

The prefactors in the above equations represent a mixture of ``primitive''
times for various computing operations as well as, for brevity, some
unimportant physical parameters. But in general we expect a performance
crossover which favors the Ewald method for low $N$ because of a smaller
overhead and eventually at large $N$, the \algo{P3M} method becomes much
faster.

\begin{figure}[tbp]
\begin{center}
\includegraphics[scale=0.5]{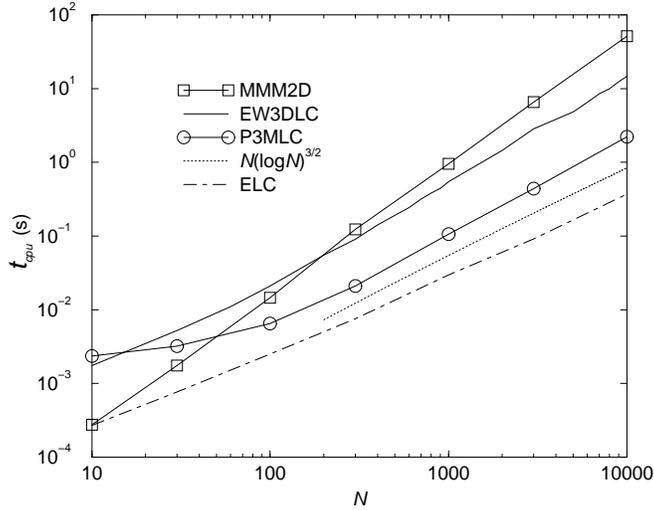}
\end{center}
\caption{ Optimal CPU times for \algo{P3MLC}, \algo{EW3DLC} and \algo{MMM2D}
  algorithms. The dotted curve displays the asymptotic scaling of the
  \algo{P3M} algorithm, and the dot-dashed curve shows the CPU time needed for
  computing the \algo{ELC}-term.}
\label{time fig}
\end{figure}

By generating random systems for a broad range of $N$ it is possible to obtain
an estimate for the crossover. We compare what are perhaps the three best
algorithms for slab-like systems in Fig. \ref{time fig} at a fixed error,
$\Delta f=10^{-2}$ $e^{2}/4\pi \epsilon L^{2}$, including the Coulomb
prefactor that has been until now omitted. This level of accuracy is as a
value that is practical for Langevin and Brownian dynamics simulations. For
instance, in a typical $N=3600$, $\rho =4.2\cdot 10^{-3}$ polyelectrolyte
simulation, the random forces are of order $2\cdot 10^{3}$ times larger than
the above RMS electrostatic force error.

The recently introduced non-Ewald \algo{MMM2D} method we recall has a
theoretical complexity $O(N^{5/3})$. The break-even point between the
\algo{MMM2D} and \algo{P3MLC} methods at this low accuracy is approximately.
$N^{MMM2D-P3MLC}=50$ and this would increase monotonically with increasing
accuracy. For reference the dotted line in Fig. \ref{time fig} indicates
$N(\log (N))^{3/2}$ scaling as is expected for the \algo{P3MLC} method. The
\algo{EW3DLC} curve plotted shows two break-even points at
$N^{MMM2D-EW3DLC}=200$ and $N^{EW3DLC-P3MLC}=17$. The shape of these curves
for the low $N$ region of the plot is dictated by the computational overhead
which can be linear or constant depending on the algorithm. It is necessary to
make Verlet lists outside of the timing loop to avoid the quadratic
determination of the pair distances in the real-space calculation.
Alternatively, these can be subtracted at the end if one accurately measure
the primitive time of this step, as we have done. Note also that the layer
correction part of the calculation used roughly 20\% of the computation time
in the scaling regime and the upper bound on its error was fixed at $10^{-2}$.

Finally we make a comparison between the speeds of \algo{EW3DLC} and
\algo{EW3DC}. Since these methods have the same scaling there is only a
difference in the prefactor. But the larger gap size required in \algo{EW3DC}
mandates an increase of the reciprocal space cutoff in order to maintain the
same accuracy. Thus there is a computational tradeoff between computing the
extra layer correction or expanding the region of reciprocal space summation.
We have used three systems with different shapes and $N=1000$ for the
comparison.  The first system has a cubic particle volume sized $1\times
1\times 1$ while the other two are pancake and cigar shaped with $h=0.5$ and
$2.0$ respectively. The parameters $(\alpha ,r_{c},k_{c},l_{c},L_{z})$ were
optimized with respect to fixed RMS errors of $3.0,4.8$ and $1.9\cdot 10^{-2}
$ for the pancake, cube and cigar respectively. Again the \algo{ELC} error was
chosen conservatively by setting the upper bound to $10^{-2}$. These values
were chosen in order to compensate for the different volumes of these systems
so that the error at the same density is the same for all three shapes.

\begin{table}[tbp]
\caption{Comparison of \algo{EW3DC} and \algo{EW3DLC}.}
\begin{tabular}{c||cc|cc}
& \algo{EW3DLC} & & \algo{EW3DC} & \\\hline \hline
system &  $(\alpha,r_{c},k_{c},L_z)$&  time (s)  &  $(\alpha,r_{c},k_{c},L_z)
$  &  time (s) \\  \hline
pancake&  $(10.20,0.271,9,0.75)$    &0.266 &$(8.30,0.321,8,2.0)$ &
 0.342  \\
cube   &  $(8.00,0.346,7,1.5)$      &0.261 &$(7.00,0.384,7,3.0)$ & 0.356  \\
cigar  &  $(6.80,0.410,6,2.4)$      &0.257 &$(6.90,0.398,7,4.0)$ & 0.368  \\
\end{tabular}
\label{tab2}
\end{table}

The results of this analysis are displayed in Table \ref{tab2}. Under these
conditions, the \algo{EW3DLC} method is only slightly (30\%) faster. The
reason is that much of the work in these calculations is performed in the
real-space and therefore the extra amount of reciprocal-space required by
\algo{EW3DC} due to the enlarged gap is not a major issue. In fact it is
almost compensated by the time used by layer correction term. For larger $N$,
\algo{EW3DLC} should become faster than \algo{EW3DC} since the
reciprocal-space time becomes much longer than the layer correction time. The
same is true about higher accuracy conditions
as is observable from the analytical error estimates (Eqs. \ref{Real_error}-%
\ref{Fourier_error}). Moreover, in a mesh implementation, the same
comparison should yield a more substantial advantage to the layer correction
since more of the computational effort is expended in reciprocal space.


\section{Discussion}

\label{disc}

We have presented and tested a method with broad applicability and low
computational cost for computing the electrostatic forces or energy of a cell
of charges with two-dimensional periodicity. This method has a complexity that
is virtually linear when applied to a mesh, so that only minor further
improvements in computation time can be attempted through a reduction of
computational overhead. The algorithm differs from \algo{EW3DC} only by the
effort to program the \algo{ELC}-term, but once this is done the errors are
easier to control and there is a greater degree of freedom in choosing the box
edges.  This makes it easy to apply to cubic versions of the \algo{P3M}
algorithm (which was one of our initial motivations). For larger systems and
better accuracy as may be required in Monte Carlo and some molecular dynamics
simulations the method should be considerably faster than \algo{EW3DC}.

It is interesting to estimate the largest system tractable with modern
computing. The \algo{P3M} algorithm can be optimized to be at least twice as
fast as the version presented here, so an $N=10000$ calculation could be
carried out in approximately. 2 seconds. A 100-node parallel computer with a
90\% scale factor could handle a system of 0.45 million particles in 1 second
-- which is sufficient for a many molecular dynamics and Monte Carlo
simulations.

Finally we can draw attention to several issues suitable for further research.
Most importantly we have not treated the case of dissimilar dielectric
materials at the slab surfaces. Because of the symmetry of the consequent
charge reflections there could be a way to handle the problem efficiently
within the current framework. Another topic is the clarification of the dipole
term for slab-wise summation for different dielectric constants at the
interface between the periodic supersystem and the surroundings.


\acknowledgments
The authors thank Zuowei Wang and Florian M\"uller-Plathe for helpful
discussions. Financial support from the DFG ``Schwerpunkt Polyelektrolyte''
is gratefully acknowledged.


\begin{appendix}
\section*{\algo{ELC}-term Forces and Pressure}

For quick reference the force and pressure formulas for the layer correction
are given here. The layer correction force on particle $i$ is 
\begin{eqnarray}
{\bf f}_{i}^{(lc)} &=&  \nonumber \\
&&-\frac{8\pi }{L_{x}L_{y}}\Biggl(\sum_{j}q_{i}q_{j}\Bigl[\;2%
\sum_{k_{x},k_{y}>0}\frac{k_{x}\cosh (k_{\parallel }z_{ij})}{k_{\parallel
}(e^{k_{\parallel }L_{z}}-1)}\sin (k_{x}x_{ij})\cos (k_{y}y_{ij})+  \nonumber
\\
&&\sum_{k_{x}>0}\frac{\cosh (k_{x}z_{ij})}{(e^{k_{x}L_{z}}-1)}\sin
(k_{x}x_{ij})\Bigr],  \nonumber \\
&&\sum_{j}q_{i}q_{j}\Bigl[\;2\sum_{k_{x},k_{y}>0}\frac{k_{y}\cosh
(k_{\parallel }z_{ij})}{k_{\parallel }(e^{k_{\parallel }L_{z}}-1)}\cos
(k_{x}x_{ij})\sin (k_{y}y_{ij})+  \nonumber \\
&&\sum_{k_{y}>0}\frac{\cosh (k_{y}z_{ij})}{(e^{k_{y}L_{z}}-1)}\sin
(k_{y}y_{ij})\Bigr],  \nonumber \\
&&-\sum_{j}q_{i}q_{j}\Bigl[\;2\sum_{k_{x},k_{y}>0}\frac{\sinh (k_{\parallel
}z_{ij})}{(e^{k_{\parallel }L_{z}}-1)}\cos (k_{x}x_{ij})\cos (k_{y}y_{ij})+ 
\nonumber \\
&&\sum_{k_{x}>0}\frac{\sinh (k_{x}z_{ij})}{(e^{k_{x}L_{z}}-1)}\cos
(k_{x}x_{ij})+\sum_{k_{y}>0}\frac{\sinh (k_{y}z_{ij})}{(e^{k_{y}L_{z}}-1)}%
\cos (k_{y}y_{ij})\Bigr]\Biggr).
\end{eqnarray}
The diagonal components of the pressure tensor can be obtained from the
partition function. For example the normal component is $P_{\bot }=P_{zz}=-%
\frac{1}{L_{x}L_{y}}\frac{\partial E}{\partial L_{z}}$. This yields 
\begin{eqnarray}
P_{\bot }^{(lc)}=\frac{8\pi ^{2}}{L_{x}^{2}L_{y}^{2}}%
\sum_{i,j=1}^{N}q_{i}q_{j}\Biggl(\;2\sum_{k_{x},k_{y}\in {\Bbb Z},>0} &&\nu
(k_{\parallel })\cos (k_{x}x_{ij})\cos (k_{y}y_{ij})+  \nonumber \\
\sum_{k_{x}>0} &&\nu (k_{x})\cos (k_{x}x_{ij})+  \label{lc P norm} \\
\sum_{k_{y}>0} &&\nu (k_{y})\cos (k_{y}y_{ij})\Biggr)
\end{eqnarray}
where 
\begin{equation}
\nu (u):=\frac{\left( \sinh uz_{ij}\right) (z_{ij}/L_{z})e^{uL_{z}}-\left(
\sinh uz_{ij}\right) (z_{ij}/L_{z})-\left( \cosh uz_{ij}\right) e^{uL_{z}}}{%
(e^{uL_{z}}-1)^{2}}\allowbreak.
\end{equation}
For one of the parallel components we find 
\begin{eqnarray}
P_{xx}^{(lc)}=-\frac{8\pi ^{2}}{L_{x}^{2}L_{y}^{2}}\sum_{i,j=1}^{N}q_{i}q_{j}%
\Biggl(\;2\sum_{k_{x},k_{y}\in {\Bbb Z},>0} &&\left[ \frac{k_{x}^{2}}{%
k_{\parallel }^{2}}\nu (k_{\parallel })+\frac{k_{y}^{2}\cosh (k_{\parallel
}z_{ij})}{k_{\parallel }^{3}L_{z}(e^{k_{\parallel }L_{z}}-1)}\right] \cos
(k_{x}x_{ij})\cos (k_{y}y_{ij})+  \nonumber \\
\sum_{k_{x}>0} &&\nu (k_{x})\cos (k_{x}x_{ij})+  \nonumber \\
\sum_{k_{y}>0} &&\frac{1}{k_{y}L_{z}}\frac{\cosh (k_{y}z_{ij})}{%
(e^{k_{y}L_{z}}-1)}\cos (k_{y}y_{ij})\Biggr).
\end{eqnarray}
Obviously this is cumbersome since after decomposing the trigonometric and
hyperbolic terms to obtain a formula that can be evaluated linearly in $N$ the
expanded version will contain over 80 terms. If rapid pressure calculations
are not demanded one can omit this term and increase $L_z$. One third of the
trace of this tensor will as usual yield the scalar pressure, $E^{(lc)}/3V$.

Another consideration in obtaining the pressures is the treatment of the
dipole term. Fortunately, a shear or distortion on the system does not result
in a distortion in the shape of the supersystem. This is in contrast to the
treatment of standard Ewald summation because the spherical summation order
requires a complex analysis of the consequences of a distorted
sphere.\cite{smith94a}

\end{appendix}

\bibliographystyle{aip}

\begin{thebibliography}{10}

\bibitem{widmann97a}
A.~H. Widmann and D.~B. Adolf,
\newblock Comp. Phys. Comm. {\bf 107}, 167 (1997).

\bibitem{parry75a}
D.~Parry,
\newblock Surf. Sci. {\bf 49}, 433 (1975).

\bibitem{parry76a}
D.~E. Parry,
\newblock Surf. Sci. {\bf 54}, 195 (1976).

\bibitem{heyes77a}
D.~M. Heyes, M.~Barber, and J.~H.~R. Clarke,
\newblock J. Chem. Soc. Faraday Trans. II {\bf 73}, 1485 (1977).

\bibitem{deleeuw82a}
S.~W. de~Leeuw and J.~W. Perram,
\newblock Physica {\bf 113A}, 546 (1982).

\bibitem{porto00a}
M.~Porto,
\newblock J. Phys. A {\bf 33}, 6211 (2000).

\bibitem{sperb01a}
R.~Strebel and R.~Sperb,
\newblock Molecular Simulation {\bf 27}, 61 (2001).

\bibitem{arnold01b}
A.~Arnold and C.~Holm,
\newblock Chem. Phys. Lett., in press, and preprint cond-mat/0202265.

\bibitem{empty-space}
Technically this space is not ``empty'' rather, it is filled with a solvent of
  the same dielectric constant.

\bibitem{spohr97a}
E.~Spohr,
\newblock J. Chem. Phys. {\bf 107}, 6342 (1997).

\bibitem{deleeuw80a}
S.~W. de~Leeuw, J.~W. Perram, and E.~R. Smith,
\newblock Proc. R. Soc. Lond. A {\bf 373}, 27 (1980).

\bibitem{deleeuw80b}
S.~W. de~Leeuw, J.~W. Perram, and E.~R. Smith,
\newblock Proc. R. Soc. Lond. A {\bf 373}, 57 (1980).

\bibitem{yeh99a}
I.-C. Yeh and M.~L. Berkowitz,
\newblock J. Chem. Phys. {\bf 111}, 3155 (1999).

\bibitem{smith81a}
E.~R. Smith,
\newblock Proc. R. Soc. Lond. A {\bf 375}, 475 (1981).

\bibitem{arnold01d}
A.~Arnold, J.~de~Joannis, and C.~Holm,
\newblock  {\bf submitted} (2002), preprint cond-mat/0202399.

\bibitem{greengard87a}
L.Greengard and V.~Rhoklin,
\newblock J. Comp. Phys. {\bf 73}, 325 (1987).

\bibitem{esselink94a}
K.~Esselink,
\newblock Comp. Phys. Comm. {\bf 87}, 375 (1995).

\bibitem{lekner91a}
J.~Lekner,
\newblock Physica A {\bf 176}, 485 (1991).

\bibitem{ewald21a}
P.~Ewald,
\newblock Ann. Phys. {\bf 64}, 253 (1921).

\bibitem{hockney88a}
R.~W. Hockney and J.~W. Eastwood,
\newblock {\em Computer Simulation Using Particles},
\newblock IOP, London, 1988.

\bibitem{sum-order}
The summation order of the layer correction is implicitely cylindrical. It
  probably does not matter whether the summation procedes along the radius of
  the cylinder first and then along its axis (as written) or vice versa. But
  they may not proceed simultaneously.

\bibitem{kolafa92a}
J.~Kolafa and J.~W. Perram,
\newblock Molecular Simulation {\bf 9}, 351 (1992).

\bibitem{petersen95a}
H.~G. Petersen,
\newblock J. Chem. Phys. {\bf 103}, 3668 (1995).

\bibitem{deserno98b}
M.~Deserno and C.~Holm,
\newblock J. Chem. Phys. {\bf 109}, 7694 (1998).

\bibitem{deserno98a}
M.~Deserno and C.~Holm,
\newblock J. Chem. Phys. {\bf 109}, 7678 (1998).

\bibitem{smith94a}
E.~R. Smith,
\newblock J. Stat. Phys. {\bf 77}, 449 (1994).

\end{thebibliography}


\end{document}